\documentclass[letterpaper, 10 pt, conference]{ieeeconf}  

\IEEEoverridecommandlockouts                              
\usepackage[utf8]{inputenc}
\usepackage[T1]{fontenc}

\usepackage{amsmath} 
\usepackage{amssymb}  
\usepackage{booktabs}
\usepackage{multirow}
\usepackage{hyperref}
\usepackage{mathtools}
\usepackage{caption}
\usepackage{subcaption}
\usepackage{bm}
\usepackage{fancyhdr}

\title{\LARGE \bf
Wellbeing in Future Mobility: Toward AV Policy Design to Increase Wellbeing through Interactions
}

\author{Shashank Mehrotra$^{1}$, Zahra Zahedi$^{1,2}$, Teruhisa Misu$^{1}$, and Kumar Akash$^{1}$%
\thanks{$^{1}$Shashank Mehrotra, Teruhisa Misu, and Kumar Akash are with Honda Research Institute USA. Inc,
        {\tt\small (shashank\_mehrotra, tmisu, kakash)@honda-ri.com}}%
\thanks{$^{2}$Zahra Zahedi is with School of Computing and AI, Arizona State University, USA. Work done during an internship at Honda Research Institute USA, Inc.
        {\tt\small (zzahedi@asu.edu)}}%
\thanks{\copyright 2023 IEEE. This is the author’s version of the work. The definitive Version of Record is published in 2023 IEEE 26th International Conference on Intelligent Transportation Systems (ITSC), September 24–28, 2023, Bilbao, Bizkaia, Spain. Personal use of this material is permitted.  Permission from IEEE must be obtained for all other uses, in any current or future media, including reprinting/republishing this material for advertising or promotional purposes, creating new collective works, for resale or redistribution to servers or lists, or reuse of any copyrighted component of this work in other works.}%
}

\DeclareMathOperator*{\argmax}{arg\,max}

\begin{document}

\maketitle

\begin{abstract}
Recent advances in Automated vehicle (AV) technology and micromobility devices promise a transformational change in the future of mobility usage. These advances also pose challenges concerning human-AV interactions. To ensure the smooth adoption of these new mobilities, it is essential to assess how past experiences and perceptions of social interactions by people may impact the interactions with AV mobility. This research identifies and estimates an individual’s wellbeing based on their actions, prior experiences, social interaction perceptions, and dyadic interactions with other road users. An online video-based user study was designed, and responses from 300 participants were collected and analyzed to investigate the impact on individual wellbeing. A machine learning model was designed to predict the change in wellbeing. An optimal policy based on the model allows informed AV actions toward its yielding behavior with other road users to enhance users' wellbeing. The findings from this study have broader implications for creating human-aware systems by creating policies that align with the individual state and contribute toward designing systems that align with an individual’s state of wellbeing.

\end{abstract}

\section{Introduction}








Recent innovations in mobility on developing autonomous vehicles \cite{sun2021shaping} and delivery robots hold promise not only toward safety but also toward comfortable and satisfactory mobility interactions \cite{jorlov2017seating}. With advances in automated vehicle (AV) technology and shared mobility \cite{stocker2017shared,hunter2023future}, there have been civic planning and engagement efforts to consider the demands at a societal level of such mobilities \cite{nigro2022investigating,mehrotra2023trust}. However, the adoption of these technologies require societal acceptance as these can potentially change traffic interactions\cite{zhang2019driving}. Specifically, these new forms of AVs should be able to coexist with existing road users.  However, there have been several societal challenges toward this smooth transition. As manifested in the recent pushback on adopting such mobility, in the 2023 Paris referendum, 89\% of voters supported a ban on electric scooters \cite{nouvian2023paris}. One of the primary reasons for such public sentiment is the perception of disregard towards public space and potential conflicts with other road users \cite{angiello2023european}. To avoid this, mechanisms are needed to increase human acceptance and satisfaction.

Primarily, researchers have focused on the safety and comfort of road users. The major promise of AV technology has been towards enhancing safety for drivers and road users \cite{gerla2014internet}. A roadblock to the  adoption of AV technology has been the challenge of public perception of their capabilities and how that would impact other road users they would interact with \cite{moody2020public}. With the future of mobility dependent on AV and humans existing in harmony, it is critical that the efficacy of these interactions is not merely judged on safety parameters alone but should consider factors that would impact the overall \emph{wellbeing} of the humans. Thus, we propose wellbeing as a critical factor for assessing mobility users' safety, satisfaction, and comfort. Amongst world institutions, wellbeing has been considered a potential proxy measure to define the state of an individual \cite{angner2010subjective}. Wellbeing could be defined as ``when individuals have the psychological,  social and physical resources to meet challenges related to those resources'' \cite{dodge2012challenge}. Studies have found better wellbeing to be associated with greater satisfaction with transportation mobility used for commuting by survey participants \cite{vella2013significance}. A systematic review of factors underpinning unsafe traffic behavior showed higher levels of wellbeing can help mitigate potential driving-related violations\cite{dorrian2022survey}. Additionally, trust in the AV and satisfaction with the interaction are critical factors that can impact users' adoption of the new mobilities. Individuals may face challenges due to unintended actions from AV mobility, which may not align with the preferences of the user \cite{sajedinia2022investigating}, as well as the perceptions of other road users about their AV mobility \cite{madigan2019understanding}.

This research aims to identify and estimate whether an individual's wellbeing is associated with the type of actions they observe and how they want their AV mobility to interact with sidewalk road users. As a first step toward addressing this challenge, this work focuses on how interactions elicit a response when they encounter a dyadic bi-directional interaction when there is symmetry in action. Specifically, we investigate accommodative actions, consisting of yielding and unyielding actions, during a conflict of the path in an interaction. An observational user study was conducted to understand the empirical evidence on whether wellbeing will impact the intention, satisfaction, and the type of action chosen by traffic participants. Based on the findings, we propose a predictive model to assess the increase and decrease in their  wellbeing based on the interaction and their inherent state of wellbeing. Through this machine learning-based approach, we propose a novel approach toward furthering the relationship between individual factors that may influence the individual's wellbeing. To our understanding, no existing research has looked to predict the state of wellbeing based on situational interactions and the current state of wellbeing.

\section{User study}

We used an online user study to evaluate the influence of interactions between a delivery robot and an AV scooter. This experiment assessed different facets of an individual's trust, wellbeing, and travel satisfaction. This section describes the details of the study.

\subsection{Materials}
\begin{figure}[t]
    \centering
    \includegraphics[width=0.8\linewidth]{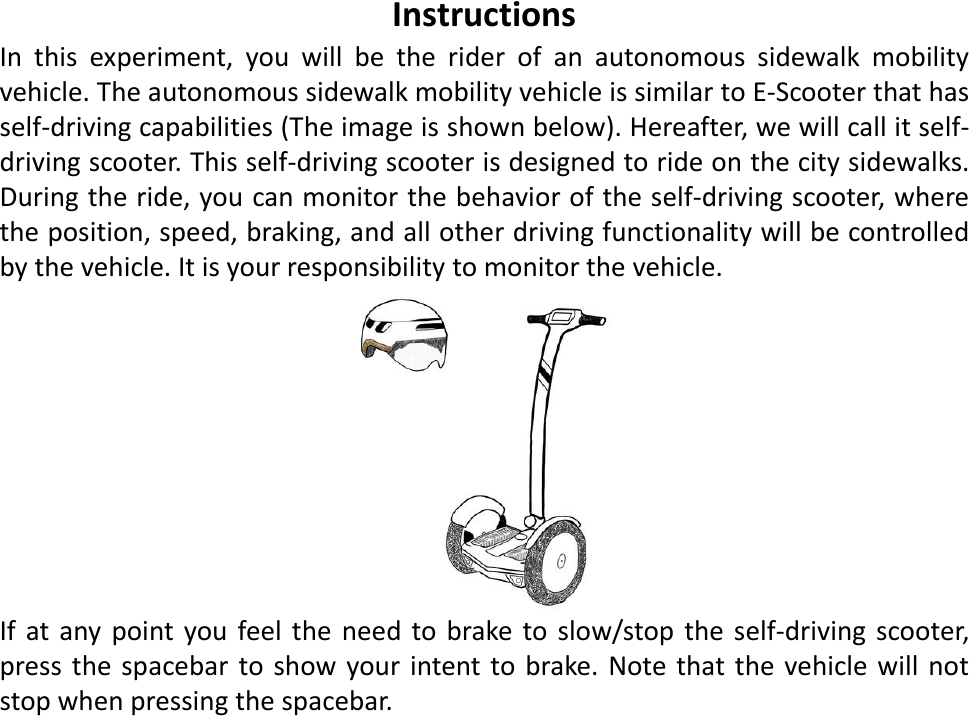}
    \caption{Information about the mobility at the beginning of the experiment.}
    \label{fig:instruction}
\end{figure}

\begin{figure}[t]
    \centering
    \includegraphics[width=0.8\linewidth]{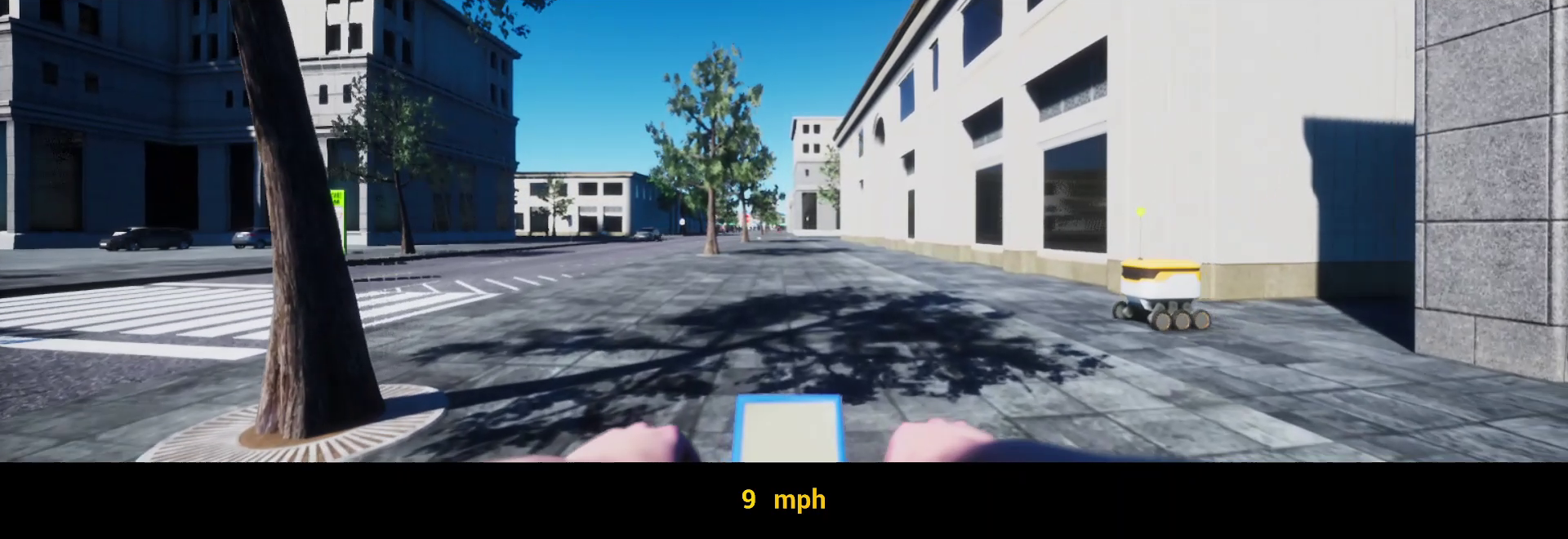}
    \caption{Layout of the scenario shown to the participants.}
    \label{fig:sample}\vspace{-1.8em}
\end{figure}

The scenarios were based on an urban-mixed traffic environment, similar to downtown areas. The participant observes the ego view of an individual riding an AV scooter. At the beginning of the experiment, participants were informed about the mobility design and how the participant is required to monitor the vehicle. The participants'  instructions and the mobility design are shown in Figure~\ref{fig:instruction}. For the online study, video recordings of a custom driving simulator were used to simulate the scenarios. The simulated environments were created using Unreal Engine 4.27 \cite{epicgame} with AirSim \cite{shah2018airsim}. The videos were recorded with the front-facing camera with a 133-degree horizontal field of view. The participants could observe the scooter handle and their hands as they rode it. A sample interaction is shown in Figure \ref{fig:sample}.

For each scenario, participants made decisions about how to interact with other road users. The decisions from the participant required coordination among road users where the outcome may not be rule-driven. In the scenarios, the road user was the delivery robot, and the participant expressed their intent for the AV scooter. The maneuver would be completed by the AV scooter, which may or may not align with the participant's intent.

\subsection{Scenarios}
\begin{figure*}[t]
    \centering
    \begin{subfigure}[b]{0.245\textwidth}
        \centering
        \includegraphics[width=\textwidth]{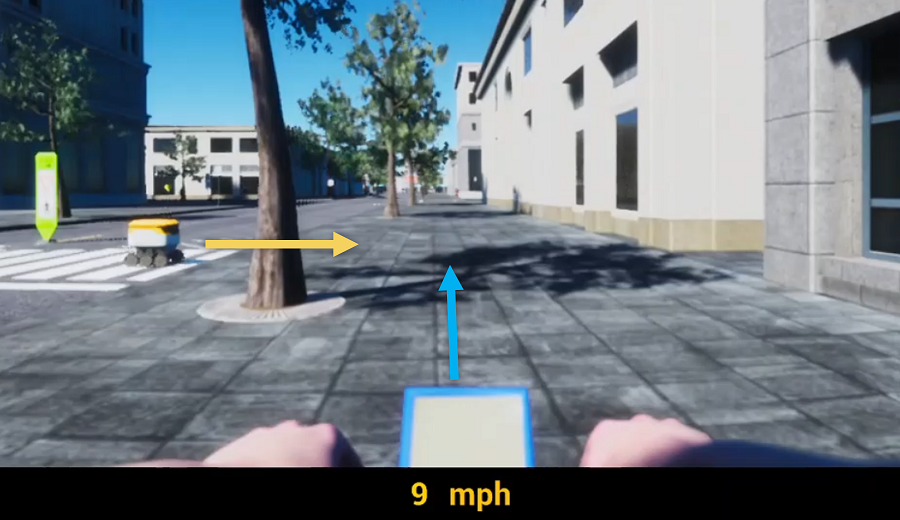}
        \caption{Scenario $S1$}\label{fig:s1}
    \end{subfigure}
    \hfill
    \begin{subfigure}[b]{0.245\textwidth}
        \centering
        \includegraphics[width=\textwidth]{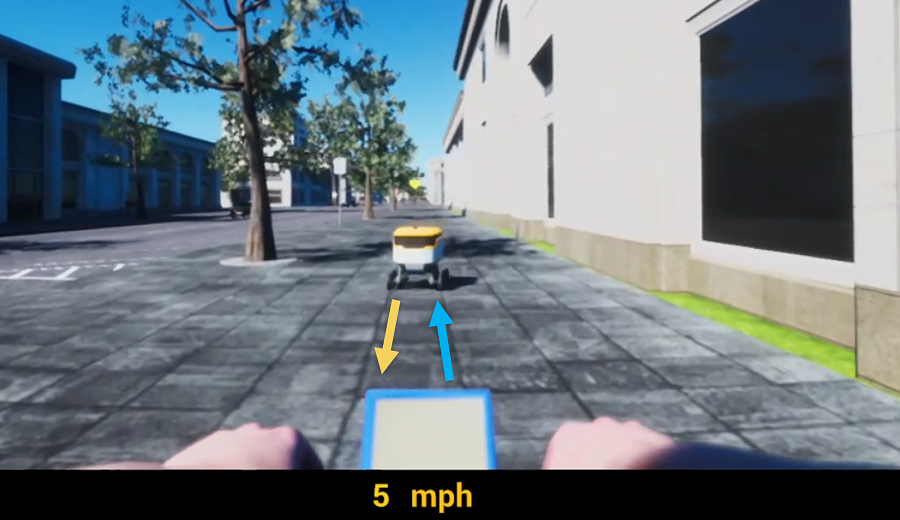}
        \caption{Scenario $S2$}\label{fig:s2}
    \end{subfigure}
    \hfill
    \begin{subfigure}[b]{0.245\textwidth}
        \centering
        \includegraphics[width=\textwidth]{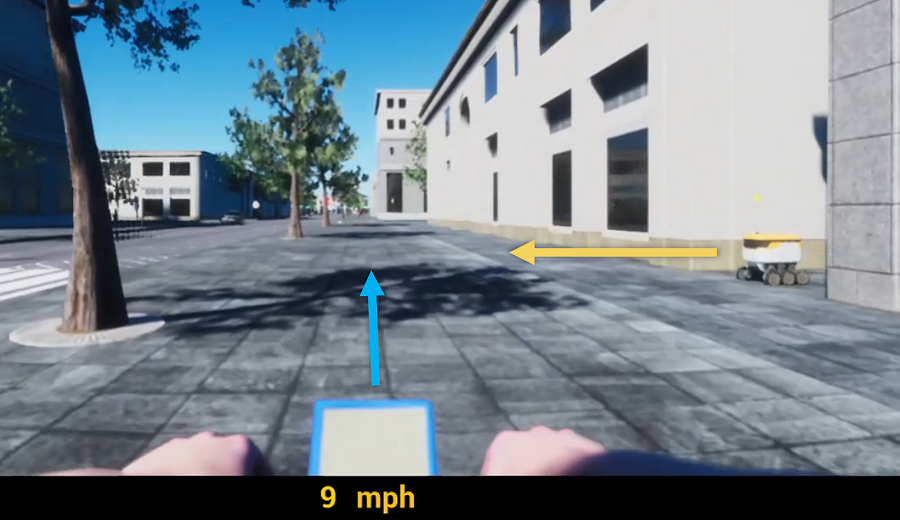}
        \caption{Scenario $S3$}\label{fig:s3}
    \end{subfigure}
    \hfill
    \begin{subfigure}[b]{0.245\textwidth}
        \centering
        \includegraphics[width=\textwidth]{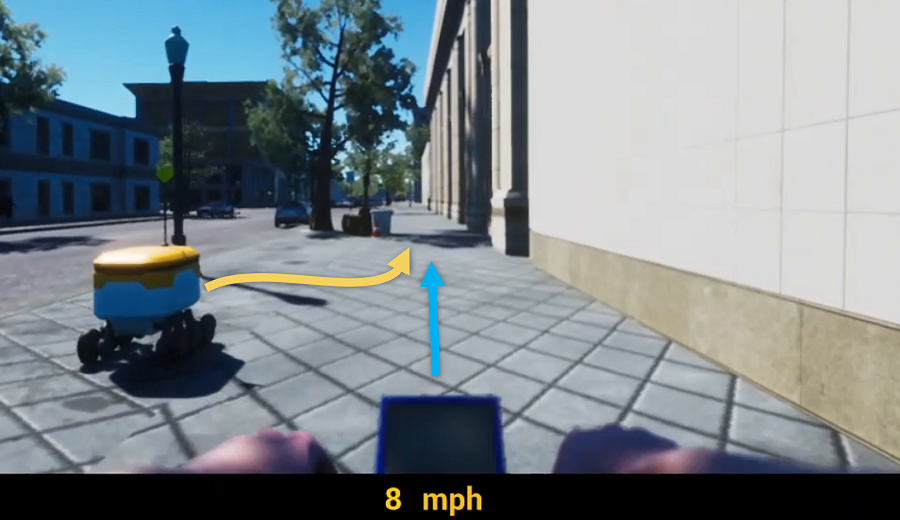}
        \caption{Scenario $S4$}\label{fig:s4}
    \end{subfigure}
    \caption{Different scenarios presented to participants. The blue arrow shows the AV scooter's path, and the yellow arrow shows the delivery robot's path.}
    \label{fig:scenario} \vspace{-1.0em}
\end{figure*}

Participants were presented with two rides, each comprising two interactions. Four scenarios were developed where the ego and sidewalk road user (delivery robot) performed either yielding or unyielding action. Figure \ref{fig:scenario} displays the four scenarios used in the study. In the first interaction, participants received yielding or unyielding behavior from the delivery robot. In the second interaction, the roles were reversed where the ego point of view performs the yielding or unyielding action, and the other road user (delivery robot) receives the action. These two interactions repeat in the next ride with different scenarios. 

\subsection{Study design}
To evaluate the impact of the interactions between the AV scooter and other road users on participants' wellbeing, a mixed design was chosen. The $2$ between-subject factors for the study design were: (1) Ego scooter's action: the AV scooter yielding to the delivery robot or unyielding to the delivery robot, and (2) Other's action: The robot yielding to the ego or unyielding to the ego. The within-subject factor for the experiment was the different scenarios presented to the participants. The levels of the scenarios were : $S2\rightarrow S1$, $S3\rightarrow S2$, $S4\rightarrow S3$, and $S1\rightarrow S4$. The scenarios were randomly chosen from a balanced Latin square combination of the $4$ scenarios. For the next ride, a pair of scenarios were chosen that were not seen in the previous interaction. The combinations of scenarios are shown in Table \ref{tab:scenario}. There were a total of 32 combinations resulting from the two 2-interaction rides, each involving four different accommodative actions. To simplify the cases, the robot's accommodative action remained the same in both rides, while the scooter could perform all possible permutations of accommodative actions across the two rides. The sequence was randomized to minimize learning effects that may compromise the study's validity. 
The accommodative behaviors in these scenarios were: (1) stopping and yielding for the delivery robot to go first($S1$ and $S3$), (2) merging ($S4$), and (3) changing their path ($S2$). Given the participant's intent, the AV scooter's behavior was aligned or non-aligned with the participant's intent to yield. Aligned means that the AV scooter would take the same action as intended by the participant. Non-aligned means the scooter AV action was different from the intention expressed by the participant. 

\subsection{Experimental conditions}
The experimental conditions were presented to the participants, where two factors - (1) robot action (yielding or unyielding, denoted as $O_y$ and $O_u$, respectively), and (2) ego AV scooter action (yielding or unyielding, denoted as $E_y$ and $E_u$, respectively) were presented as a combination to participants.
\begin{table}[t]
\centering
\caption{Order of scenarios in each interaction and accommodative action combinations}
\label{tab:scenario}
\resizebox{.45\textwidth}{!}{\begin{tabular}{cc|c}
\toprule
\multicolumn{2}{c|}{Interactions}& Accommodative actions\\
\hline
 1\textsuperscript{st} two-interaction ride& 2\textsuperscript{nd} 
two-interaction ride&\\
 $S2\rightarrow S1$& $S4\rightarrow S3$&$O_uE_u$\\
 $S3\rightarrow S2$& $S1\rightarrow S4$&$O_uE_y$\\
 $S4\rightarrow S3$& $S2\rightarrow S1$&$O_yE_u$\\
 $S1\rightarrow S4$& $S3\rightarrow S2$&$O_yE_y$\\
\bottomrule 
\end{tabular}}
\end{table}

\subsection{Study Procedure}
\subsubsection{Participant recruitment} 
A total of $300$ participants (161 males, 132 females, and 7 others; mean age of 39.35, SD = 13.55) were recruited from the Prolific\footnote{Prolific Academic Ltd. (\url{www.prolific.co})} online study platform. Participants were required to be adults (greater than 18 years old) holding a valid US driver's license. For their participation, each participant received $\$3.0$ as compensation for completing the survey, which took approximately $25$ minutes to finish. All participants provided informed consent, and the study was approved by the Bioethics Committee in Honda R\&D (approval code: 99HM-065H).

\begin{figure*}[t]
    \centering
    \includegraphics[width=\textwidth]{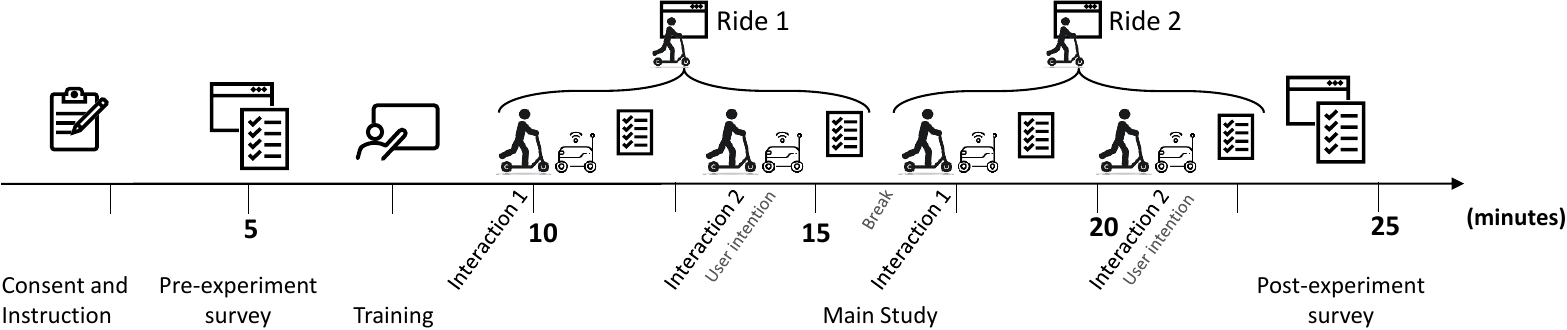}
    \caption{The procedure followed by all the participants}
    \label{fig:procedure}\vspace{-1.5em}
\end{figure*}

\subsubsection{Procedure} 
Participants received instructions and completed a pre-experiment survey to measure their initial social interaction perceptions that captures their content, positive relationship, and wellbeing (see Table~\ref{tab:socialwellbeing}). Participants then underwent a 3-minute training session on the web-based driving simulation. The training included familiarization with the simulation, other road users, and how participants interact with delivery robots. They were also instructed to complete the survey post-interactions. All the survey questions were displayed in a pop-up window, similar to the main study. Participants were instructed on how the scooter communicates the situation in a descriptive voice to the user and how to indicate their intent to brake or decelerate using the space bar. Upon completing the tutorial, each participant completed two rides consisting of two interactions. After each interaction, participants answered questions about their wellbeing, trust, positive relationship, and satisfaction. In the second interaction, they were asked to indicate their preferred action for their self-driving scooter before they saw the scooter's behavior toward the robot. After completing both rides, participants answered questions about their demographic information and previous experience with autonomous features in vehicles, and prior micromobility use. In the end, they were compensated for their participation. The entire experimental procedure is illustrated in Figure \ref{fig:procedure}

\begin{table}[t]
    \centering
     \caption{Initial questionnaire to capture social interaction perceptions.}
    \resizebox{\columnwidth}{!}{%
    \begin{tabular}{p{9cm}l}
    \toprule
    \textbf{1. Social wellbeing: Content - } Generally, I am content with the relation with sidewalk and road users\\
    \textbf{2. Social wellbeing: Positive relation - } Generally, I think the sidewalk and road users around me interact with each other in a positive manner\\
    \textbf{3. Social wellbeing: Ego wellbeing - } Generally, experience with sidewalk and road users contributes to my wellbeing\\
    \bottomrule 
    \end{tabular}}
    \vspace{-8pt}
    \label{tab:socialwellbeing}
\end{table}

\subsubsection{Attention Checks and Commitment Question}
To ensure high data quality from attentive and motivated participants in the online study, five attention check questions and a commitment request were included in the survey. Studies show that asking respondents to commit to providing thoughtful responses decreases the rate of quality issues \cite{qualtricsUsingAttention}. Post-training,  participants were asked to commit to providing thoughtful answers. For the first attention check question, a counterfactual statement was presented to assess the attention of participants (e.g., ``I work fourteen months in a year'': yes or no) in the pre-experiment survey. This statement was incorporated to elicit the correct response from attentive respondents \cite{huang2015detecting}. The main survey included four additional questions (two per ride) consisting of instructed response items and special attention checks. Specifically, respondents were asked to select a specific response category (e.g., select ``strongly disagree'') \cite{silber2022issue}. No data was recorded for the participants who failed two or more attention checks. 

\subsection{Measurements}

\begin{table}[t]
    \centering
     \caption{Wellbeing (1--7) and Trust (8) questionnaire}
    \resizebox{\columnwidth}{!}{%
    \begin{tabular}{p{10cm}l}
    \toprule
    \textbf{1. Positive relationship (me to others):} Based on the current interaction, I am content with the relation with other robot.\\
    \textbf{2.	Positive relationship (others to me):} Based on the current interaction, I think the delivery robot around me handle others in a positive manner.\\
    \textbf{3.	Satisfaction (positive activation):} During my current travel event I w;as worried/confident.\\
    \textbf{4.	Satisfaction (positive deactivation):} During my current travel event I was tired/alert.\\
    \textbf{5.	Satisfaction (cognitive evaluation):} My current travel event worked poorly/worked well.\\
    \textbf{6.	Wellbeing:} This travel event contributes to my wellbeing.\\
    \textbf{7.	Trust (in others):} Based on the current interaction, I trust robots in my surrounding.\\
    \textbf{8.	Trust (in scooter):} Based on the current interaction, I trust my self-driving scooter.\\
    \bottomrule 
    \end{tabular}}
    \vspace{-8pt}
    \label{tab:questionnaire}
\end{table}


\subsubsection{Individual state of trust, wellbeing, and positive relations} 
We utilized a modified version of the social interaction perception questionnaire developed by Radzyk \cite{radzyk2014validation} to assess users' wellbeing (Q1--Q7 in Table~\ref{tab:questionnaire}). The adapted questionnaire was situational and more relevant to our study scenario. The questionnaire was designed to measure four factors: (1) positive relationship, (2) satisfaction with travel \cite{friman2013psychometric}, (3) general wellbeing, and (4) trust.  We included two questions to assess positive relationships about (1) Participants' relationships toward others (Q1) and (2) Others' relationships toward them (Q2). To measure satisfaction with travel, we included one question from the three factors in Satisfaction with Travel Scale \cite{friman2013psychometric}: positive activation (Q3), positive deactivation (Q4), and cognitive evaluation (Q5). After each interaction, we included a question to measure the user's overall sense of wellbeing (Q6). Trust was measured using a question about the user's trust in others (Q7). Additionally, Participants' trust in the self-driving scooter was also assessed independently (Q8) (see Table \ref{tab:questionnaire} for more detail). The measures of positive relations and  satisfaction with travel were the mean scores of the two responses each participant provided after each interaction.

\subsubsection{Social interaction perceptions between road users} 
Participants responded to three questions to assess the initial perception of social interactions between participants and other road users. The questions covered: (1) general contentment between sidewalk and road users,  (2) positive relationship between sidewalk and road users, and (3) perceived ego wellbeing as a result of their experience with sidewalk and road users. The statements are shown in Table \ref{tab:socialwellbeing}. The participants responded on a scale of 1 to 7, where 1 corresponds to low, whereas 7 corresponds to high.

\subsubsection{User's yielding/unyielding intention} 
To determine the user's intention in the interaction where the ego is the contributor, participants were asked: ``What action would you like your self-driving scooter to take regarding the delivery robot?'' Two options are given, one implying yielding action and the other unyielding action. By asking this question, we can assess whether the user's intention aligns with the behavior of the self-driving scooter in that particular scenario.

\subsubsection{User's braking behavior} 
To determine the user's braking behavior, key presses from the online study were recorded and analyzed. Participants expressed their intent to brake by pressing the Spacebar key on their computers to record if they felt their mobility should brake or decelerate. A gray indicator outlining the screen would show up to provide visual feedback. The scooter's actual driving behaviors are not affected by participants' button-pressing inputs. The frequency (number of brake presses) and the intensity (longest continuous brake press) were recorded to account for the braking behavior of the participants.

\begin{table*}[ht]
\setlength\tabcolsep{2.5pt}
\centering
\caption{Estimates from the regression analysis. Only estimates with p-value $< 0.05$ have been reported. *$p<0.05$; **$p<0.01$; ***$p<0.001$ }
\begin{tabular}{llrrrrr}
Dependent   Variable                      & Independent Variable                                   & Estimate  & Std. Error & $t$-value & $p$-value & $\eta_{p}^{2}$-value \\
\toprule
\multirow{4}{*}{Wellbeing\textsubscript{interaction 1}} & Robot {yielding}                                   & $0.2934$  & $0.1479$   & $1.984$   & $0.0482^{*}\hspace{3.5pt}\hspace{3.5pt}$ & $0.01$   \\
                                          & Brake frequency: Interaction 1                       & $-0.0950$ & $0.0304$   & $-3.129$  & $0.0018^{**}\hspace{3.5pt}$ & $0.02$                \\
                                          & Social interaction perception: Ego wellbeing                               & $0.4488$   & $0.0707$   & $6.344$   & $<0.0001^{***}$   & $0.12$            \\
                                          & Scenario: $S2\rightarrow S1$                               & $0.3418$  & $0.1013$   & $3.375$   & $0.0008^{***}$ & $0.04$             \\
\midrule
{Wellbeing\textsubscript{interaction 2}}       & Social interaction perception: Ego wellbeing                                    & $0.4553$  & $0.0715$   & $6.365$   & $<0.0001^{***}$ & $0.12$             \\
\midrule
\multirow{6}{*}{Satisfaction (positive activation)\textsubscript{interaction 1}}    & Robot\_yielding                            & $0.3737$    & $0.1149$     & $3.254$   & $0.0013^{**}\hspace{3.5pt}$  & $0.04$          \\
                                          & Brake frequency: Interaction 1                        & $-0.0510$    & $0.0236$    & $-2.160$   & $0.0312^{*}\hspace{3.5pt}\hspace{3.5pt}$     & $0.000812$      \\
                                          & Social interaction perception: Ego wellbeing                                & $0.1089$    & $0.0549$    & $1.983$   & $0.0484^{*}\hspace{3.5pt}\hspace{3.5pt}$      & $0.01$           \\
                                          & Scenario: $S1\rightarrow S4$                               & $0.4690$     & $0.1282$     & $3.658$   & $0.0003^{***}$  & $0.03$             \\
                                          & Scenario: $S2\rightarrow S1$                               & $0.6573$    & $0.0788$    & $8.340$    & $<0.0001^{***}$  & $0.19$            \\
                                          & Scenario: $S3\rightarrow S2$                               & $0.4052$    & $0.1269$     & $3.192$   & $0.0015^{**}\hspace{3.5pt}$  & $0.02$                \\
\midrule
\multirow{4}{*}{Satisfaction (positive activation)\textsubscript{interaction 2}}    & Robot yielding                            & $0.3665$  & $0.1154$     & $3.176$   & $0.0017^{**}\hspace{3.5pt}$ & $0.03$               \\
                                          & Scenario: $S1\rightarrow S4$                               & $0.4192$  & $0.1264$   & $3.320$    & $0.0010^{***}$  & $0.03$             \\
                                          & Scenario: $S2\rightarrow S1$                               & $0.6771$  & $0.0805$   & $8.413$   & $<0.0001^{***}$  & $0.19$             \\
                                          & Scenario: $S3\rightarrow S2$                               & $0.4144$   & $0.1276$   & $3.247$   & $0.0013^{**}\hspace{3.5pt}$ & $0.02$               \\
\midrule
\multirow{6}{*}{Trust (in others)\textsubscript{interaction 1}}          & Robot yielding                             & $0.5802$  & $0.1706$   & $3.401$   & $0.0008^{***}$  & $0.20$           \\
                                          & Brake frequency: Interaction 1                      & $-0.0964$  & $0.0389$    & $-2.470$   & $0.0138^{*}\hspace{3.5pt}\hspace{3.5pt}$ & $0.000245$                \\
                                          & Social interaction perception: Ego wellbeing                               & $0.3209$  & $0.0814$    & $3.945$   & $0.0001^{***}$ & $0.03$             \\
                                          & AV scooter yielding                                     & $-0.2429$ & $0.1138$   & $-2.133$  & $0.0334^{*}\hspace{3.5pt}\hspace{3.5pt}$  & $0.000551$              \\
                                          & Alignment                                               & $0.2950$  & $0.1135$   & $2.599$   & $0.0096^{**}\hspace{3.5pt}$ & $0.00000343$               \\
                                          & Scenario: $S1\rightarrow S4$                               & $-0.4357$ & $0.1918$   & $-2.271$  & $0.0236^{*}\hspace{3.5pt}\hspace{3.5pt}$  &$0.10$               \\
\midrule
\multirow{3}{*}{Trust (in scooter)\textsubscript{interaction 1}} & Social interaction perception: Positive relation                               & $0.2224$   & $0.0932$    & $2.385$   & $0.0177^{*}\hspace{3.5pt}\hspace{3.5pt}$ &$0.05$             \\
                                         & Social interaction perception: Ego wellbeing                                & $0.1984$   & $0.0733$    & $2.707$   & $0.0072^{**}\hspace{3.5pt}$  & $0.10$            \\
                                          & Scenario: $S2\rightarrow S1$                               & $0.6539$   & $0.1107$     & $5.907$   & $<0.0001^{***}$ & $0.10$ \\
\midrule
\multirow{3}{*}{Trust (in scooter)\textsubscript{interaction 2}}& Brake frequency: Interaction 1                    & $-0.1137$  & $0.0373$    & $-3.052$  & $0.0024^{**}\hspace{3.5pt}$ &$0.02$              \\
                                          & Social interaction perception: Ego wellbeing                               & $0.2169$   & $0.0807$     & $2.688$   & $0.0076^{**}\hspace{3.5pt}$  &$0.02$               \\
                                          & AV scooter yielding                                     & $0.7437$   & $0.1083$    & $6.864$   & $<0.0001^{***}$   & $0.10$            \\
\bottomrule
\end{tabular}
\label{tab:reganalysis}
\end{table*}

\section{Regression analysis}
Of the 300 participants, 299 were considered for the analysis (responses from 1 participant were not recorded). 
Out of all participants, only 43 participants reported using an Electric bike, 24 had used a Segway, and 76 had used an electric scooter before the experiment. Participants aged 18 to 25 reported the lowest positive experience while using micromobility, whereas those aged 36-45 reported the highest positive experience. 
\subsection{Results from the regression analysis}
A linear mixed effects model using \emph{lme4} \cite{kuznetsova2015package} package in R was used to measure the influence on trust, wellbeing, and satisfaction. For the analysis, the following model was used for the dependent variables (DVs):
\begin{align} \begin{split}
    \text{DV} \sim &\text{ Robot yielding} + \text{AV scooter yielding } +\\
    &\text{ Braking frequency} + \text{Social interaction perception } + \\
    &\text{ Scenarios} + \text{Alignment} + (1|\text{Participant}) + \epsilon  \nonumber
\end{split}\end{align}
The dependent measures were based on the survey statements reported in Table \ref{tab:questionnaire}: (1) Wellbeing (question 6), (2) Satisfaction (mean of questions 3,4 and 5), Trust (in other) (question 7), and Trust (in scooter) (question 8). Table \ref{tab:reganalysis} shows the statistically significant findings from the study ($p$-value < $0.05$). The results from the analysis found that interaction with other road users influenced self-reported measures for participants. The delivery robot's yielding behavior by the delivery robot \emph{positively} influenced the state of wellbeing, satisfaction, and trust in other road users when the participants interacted with the robot during the first interaction. 

\subsubsection{Wellbeing} The analysis found that yielding behavior from robots found an \emph{increase} in wellbeing in the first interaction. Additionally, the frequency of the intention to brake was \emph{negatively} associated with wellbeing. Out of the scenario combinations, $S2 \rightarrow S1$ was found to increase wellbeing. The Social interaction perceptions on ego wellbeing was found to \emph{increase} wellbeing for both interactions.

\subsubsection{Satisfaction} The analysis found that yielding behavior from robots found an \emph{increase} in satisfaction in both interactions. The frequency of intention to brake was \emph{negatively} associated with satisfaction in the first interaction. The perceived ego wellbeing was found to \emph{increase} satisfaction for only the first interaction. For both interactions, scenarios: $S2 \rightarrow S1$, $S3 \rightarrow S2$, and $S1 \rightarrow S4$ was found to increase satisfaction.

\subsubsection{Trust} The analysis found that yielding behavior from robots found an \emph{increase} in trust in other road users in the first interaction. The frequency of intention to brake was \emph{negatively} associated with trust in other road users in the first interaction. The perceived ego wellbeing was found to \emph{increase} trust in others for both interactions. For the first interaction, yielding behavior from the AV scooter resulted in \emph{negative} impact on trust in others, whereas the alignment in AV action and Robot action resulted in \emph{positive} influence on trust in others. The perceived ego wellbeing was found to \emph{increase} trust in others for only the first interaction. trust in AV scooter \emph{increased} due to perceived ego wellbeing, and perceived positive relations with other road users in the first interaction. In the second interaction, braking frequency \emph{decreased} trust in AV scooter, whereas yielding behavior from AV scooter \emph{increased} trust in AV scooter.  
\subsection{Implications from the findings}
Findings from the regression analysis show that the yielding behavior of the delivery robot and the yielding behavior of the AV scooter increased wellbeing and trust in the AV scooter, respectively. This implies that any yielding behavior from the delivery robot enhanced participants' wellbeing and trust. However, AV scooter yielding resulted in lower trust in other road users. This result could be attributed to the interplay between trust in AV versus trust in society, given the action of other road users. 

Results found wellbeing, satisfaction, and trust were  \emph{positively influenced} by initial social interaction perceptions. This could be attributed to an individual's initial perceptions about their social interactions and how they impact trust and wellbeing during interactions with other road users. Additionally, different scenarios also impacted participants' wellbeing and satisfaction. All scenarios except for $S1\rightarrow S4$ reported an increase in wellbeing, trust, and satisfaction. $S1\rightarrow S4$ involved the delivery robot approaching the participant from behind in $S4$, which may help explain trust and satisfaction. Results also found a \emph{negative impact} of braking frequency on trust, wellbeing, and satisfaction. This finding could be attributed to participants expressing higher perceived risk through braking. Findings from the regression analysis provide a unique perspective on how social interactions with different behaviors impact the user's state, one of which is their state of wellbeing. 

As a result, for a human-aware system to enhance users' wellbeing, the system needs to consider four critical elements while making its decisions: 1) the system should adapt its behavior based on the scenario depending on users' perception of the scenario; 2) the system should consider objective indicators such as users' braking intentions; 3) prior knowledge of users' wellbeing can assist in ensuring optimal action; and 4) the system actions should be personalized to the user based on user's perception about their social interactions.
To demonstrate this concept, we develop a preliminary model to predict the change in wellbeing by utilizing the findings from regression and propose the optimal policy to increase users' wellbeing. 


\section{Predictive Modeling and Optimal Policy}
Based on the findings from the regression analysis, we infer that the participants' wellbeing is influenced by the yielding behaviors of the robot and AV scooter, scenario type, influence of social interaction on ego wellbeing, and braking behavior from the participant. Therefore, a human-aware system that can utilize these relations to make better-informed decisions can help enhance user's wellbeing. The system must first predict the user's wellbeing and then utilize it to calculate the optimal action. To achieve this design, we propose a classification model that can predict change in wellbeing as a binary variable (decrease or increase). Hereafter, wellbeing is calculated as the average value of the user's response to the wellbeing questionnaire (Q1-7 in Table~\ref{tab:questionnaire}) to combine the effect of positive relation, satisfaction, and trust in others. 

\subsection{Model Training and Evaluation}
We consider the participants' past experience, robot and scooter actions, scenario type, and user's initial wellbeing and braking behavior as potential predictors. Participants' past experience includes their prior micromobility use and their responses to the social interaction perception questionnaires. Robot action is considered a binary variable, where 1 represents yielding behavior, and 0 represents unyielding behavior in the first interaction of a ride. Similarly, Scooter's action is also considered a binary variable with the same coding applied to the robot's action for the second interaction of a ride. The initial wellbeing is calculated based on the user's response in the first interaction. The change in wellbeing is defined as decreasing if the wellbeing response in the second interaction decreases compared to that in the first interaction. It is defined as increasing if it remains the same or increases in the second interaction. Scenario type consists of the four scenario types ($S1$-$4$) and is one-hot encoded. Finally, the user's braking behavior is defined as a binary variable of whether the user indicated to brake before the intention question in the second interaction or not. Note that we do not use alignment or user's intention as a predictor as they cannot be obtained in real applications. For the 299 participants' data, since each participant completed two rides, we have a total of 598 samples. 

Given the modest sample size, we train simple classification models to predict wellbeing change. We consider logistic regression (LR), support vector classifier (SVC) with radial basis function (RBF) kernel, random forest (RF) classifier, and AdaBoost classifier. To identify the best set of features for each of the classifiers, we use backward sequential feature selection to obtain the smallest feature set that maximizes the area under the curve (AUC) for the receiver's operating characteristics (ROC). The AUC-ROC is calculated using 5-fold cross-validation during the feature selection. Based on the selected feature set for each model, we then calculate the model's performance using 5-fold cross-validation, where each fold consists of a different set of participants. We iterate the cross-validation 1000 times with a randomized split of participants across the folds in each iteration to obtain robust performance metrics for the participants' distribution. The models are trained using the sklearn package (version 1.21) in Python with their default parameters.


Table~\ref{tab:model_perf} shows the average performance metrics of the models across the 1000 iterations with 95\% confidence interval (CI). 
The SVC outperforms other models with an average AUC-ROC of 0.7136. The best feature set selected for the SVC consists of 8 features:  
1) Prior micromobility use (yes or no), 2) Social interaction perception-Content, 3) Social interaction perception-Ego wellbeing, 4) Scenario type, 5) User's initial wellbeing, 6) Robot's action, 7) User's braking behavior, and 8) Scooter's action. 

\begin{table}
\centering
\caption{Average 5-fold cross-validation performance of the models with 95\% CI.}
\label{tab:model_perf}
\resizebox{\linewidth}{!}{%
\begin{tabular}{llll} 
\toprule
 & Accuracy & F1-score & AUC-ROC \\ 
\midrule
SVC & $69.47\% \pm 0.06\%$ & $0.7781 \pm 0.0005$ & $0.7136 \pm 0.0006$ \\
LR & $65.86\% \pm 0.05\%$ & $0.7539 \pm 0.0004$ & $0.6866 \pm 0.0004$ \\
RF & $62.52\% \pm 0.08\%$ & $0.7121 \pm 0.0007$ & $0.6275 \pm 0.0008$ \\
AdaBoost & $64.73\% \pm 0.07\%$ & $0.7345 \pm 0.0005$ & $0.6840 \pm 0.0006$ \\
\bottomrule
\end{tabular}
}
\end{table}


\subsection{Optimal Policy Design}
\begin{figure*}[t]
    \centering
    \begin{subfigure}[b]{0.3\textwidth}
        \centering
        \includegraphics[width=\textwidth]{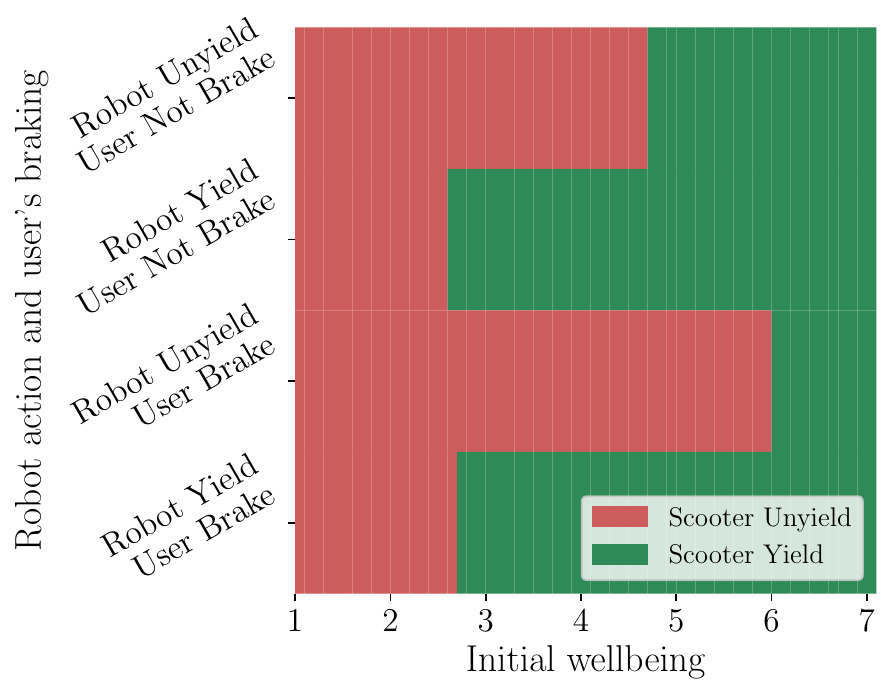}
        \caption{\footnotesize{Social interaction perception: Content $= 2$}}\label{fig:policy_socContent2}
    \end{subfigure}
    \hspace{8pt}
    \begin{subfigure}[b]{0.3\textwidth}
        \centering
        \includegraphics[width=\textwidth]{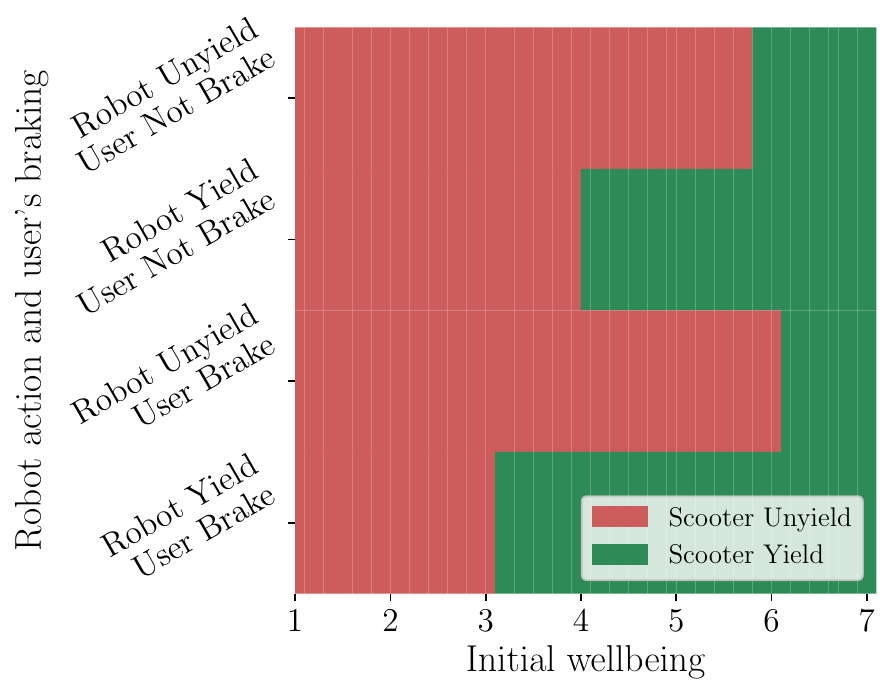}
        \caption{\footnotesize{Social interaction perception: Content $= 4$}}\label{fig:policy_socContent4}
    \end{subfigure}
    \hspace{8pt}
    \begin{subfigure}[b]{0.3\textwidth}
        \centering
        \includegraphics[width=\textwidth]{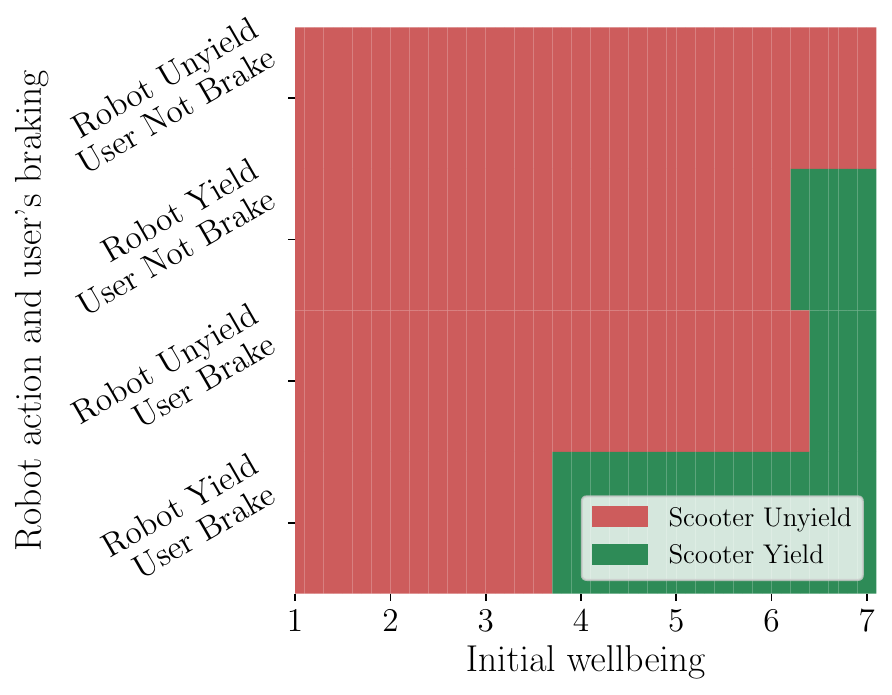}
        \caption{\footnotesize{Social interaction perception: Content $= 6$}}\label{fig:policy_socContent6}
    \end{subfigure}
    \caption{Optimal scooter action with variations with user's response to Social interaction perception: Content.  Red and green colors show the optimal action given the user's initial wellbeing, the robot's action in the prior interaction, and the user's braking behavior. }
    \label{fig:policy_socContent} 
\end{figure*}

Although the model allows us to predict users' change in wellbeing, it also helps to identify the optimal choice for the scooter's action, given the other predictors. This can be achieved as follows. Let $\bm{x}$ denote the vector of the eight predictors, $\bm{x}^{\setminus\{e\}}$ denote the vector of the seven predictors excluding the ego scooter's action, and $e\in\{E_u,E_y\}$ denote the ego scooter's action. The learned model $f: \bm{x} \rightarrow \Delta w$ predicts the likelihood of change in wellbeing $\Delta w \in \{\Delta w_\downarrow,\Delta w_\uparrow\}$ as
$$Pr(\Delta w |\bm{x}) = Pr(\Delta w |\bm{x}^{\setminus\{e\}}, e) \hspace{2pt}. $$
For a human-aware system aimed at increasing users' wellbeing, we can define the optimal policy as maximizing the likelihood of an increase in wellbeing $\Delta w_\uparrow$. Therefore, the optimal policy $\pi(e)$ for ego scooter's action $e\in\{E_u,E_y\}$ is given as
$$\pi(e) = \argmax_{e\in\{E_u,E_y\}} Pr(\Delta w_\uparrow |\bm{x}^{\setminus\{e\}}, e) \hspace{2pt}.$$
Before any interaction, if the system knows the user's past experience, social interaction perceptions, and the scenario and knows the user's braking behavior, the scooter can take the optimal yielding action.

\begin{figure*}[t]
    \centering
    \begin{subfigure}[b]{0.3\textwidth}
        \centering
        \includegraphics[width=\textwidth]{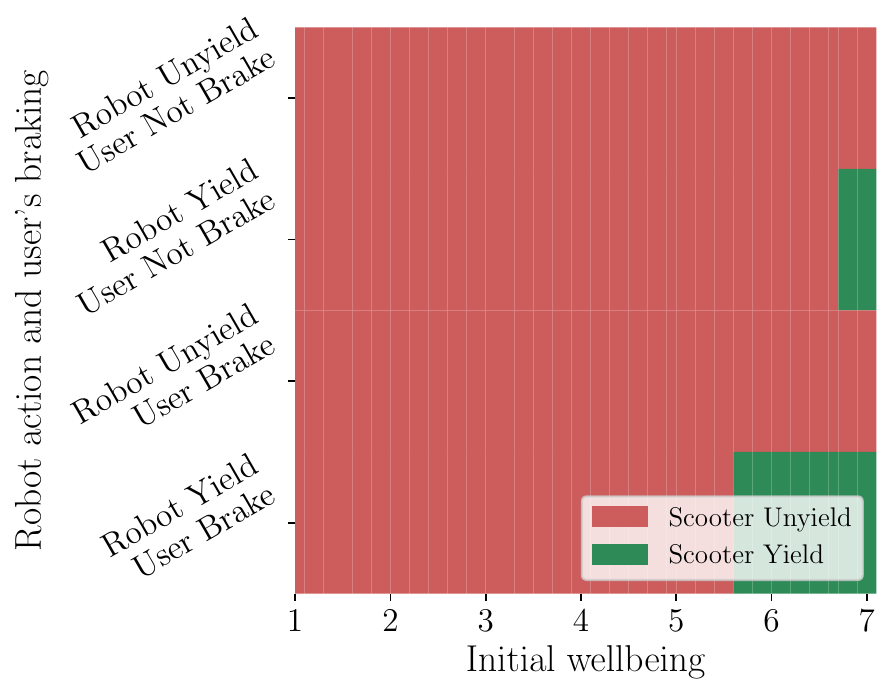}
        \caption{\footnotesize{Social interaction perception: Ego wellbeing $= 2$}}\label{fig:policy_socEgoWb2}
    \end{subfigure}
    \hspace{8pt}
    \begin{subfigure}[b]{0.3\textwidth}
        \centering
        \includegraphics[width=\textwidth]{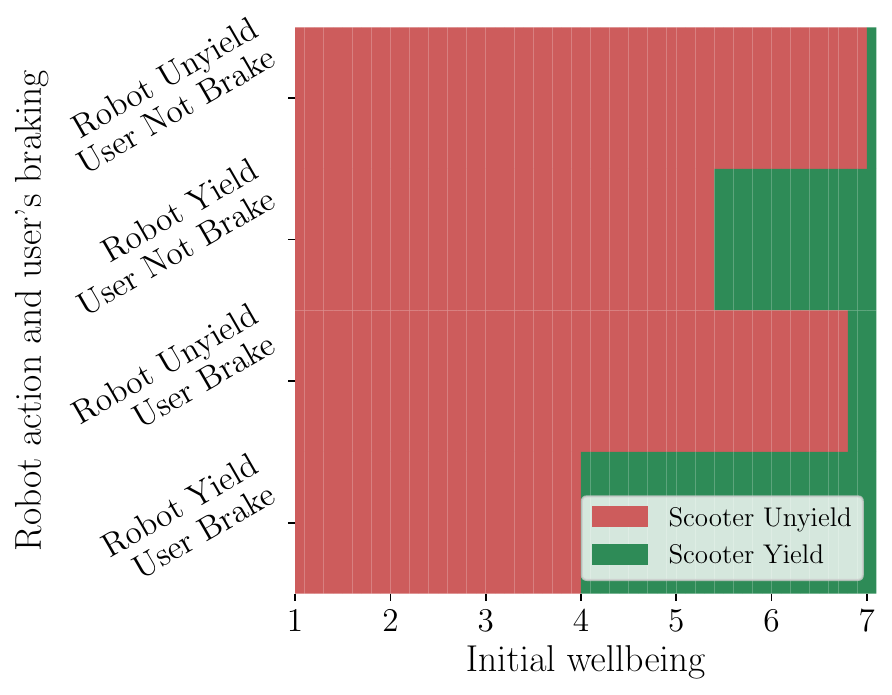}
        \caption{\footnotesize{Social interaction perception: Ego wellbeing $= 4$}}\label{fig:policy_socEgoWb4}
    \end{subfigure}
    \hspace{8pt}
    \begin{subfigure}[b]{0.3\textwidth}
        \centering
        \includegraphics[width=\textwidth]{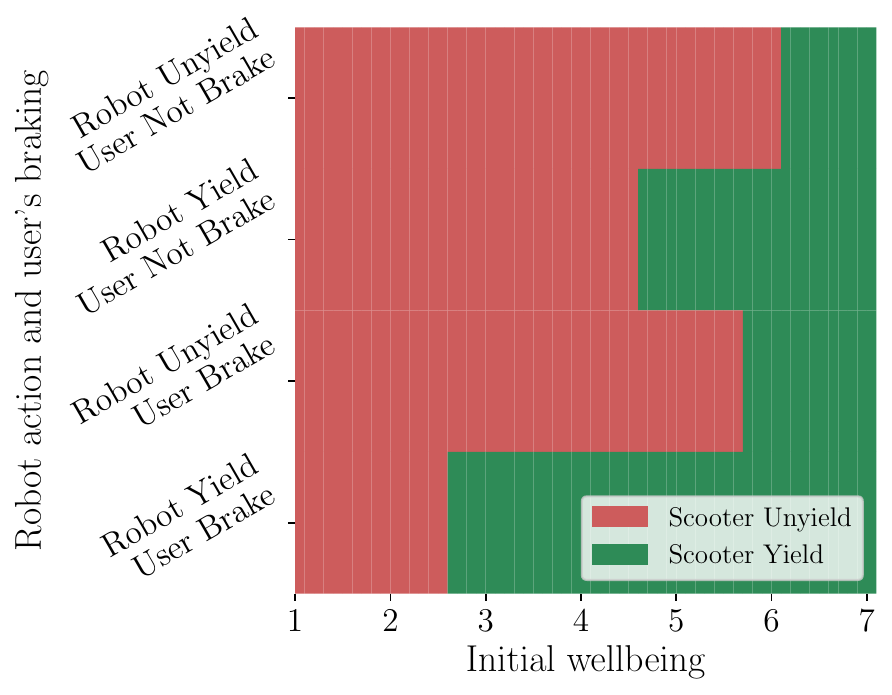}
        \caption{\footnotesize{Social interaction perception: Ego wellbeing $= 6$}}\label{fig:policy_socEgoWb6}
    \end{subfigure}
    \caption{Optimal scooter action with variations with user's response to Social interaction perception: Ego wellbeing. Red and green colors show the optimal action given the user's initial wellbeing, the robot's action in the prior interaction, and the user's braking behavior. }
    \label{fig:policy_socEgoWb} \vspace{-1.0em}
\end{figure*}

To further analyze the policy, we train the SVC model using the entire data and visualize the policy for different combinations of the other predictors. Specifically, we show the effect of an individual's social interaction perception responses and prior micromobility use affect the learned optimal policy. Figure~\ref{fig:policy_socContent} visualizes the optimal policy with varying values of social interaction perception-Content. The x-axis maps the initial wellbeing, and the y-axis shows a robot's action and user's braking behavior. The other predictors are fixed to be their median values (Prior micromobility use: No, Social interaction perception-Ego wellbeing: 5, Scenario: $S4\rightarrow S3$). The red and the green color shows the optimal scooter's action being unyielding and yielding, respectively, based on the learned policy. 

From Figure~\ref{fig:policy_socContent2}, we see that the scooter's optimal action is unyielding when the user's initial wellbeing is low and vice-versa. This is expected as individuals with lower wellbeing prioritize their needs over others and therefore do not yield to the robot. Furthermore, we observe that when the robot is unyielding to the user in the previous interaction, the learned policy shows that the user's wellbeing will increase if the AV scooter does not yield to the robot even for relatively higher levels of wellbeing. In contrast, the policy recommends that the AV scooter yielding to the robot will increase the user's wellbeing when the robot yielded in the previous interaction when the user's wellbeing is not too low. This captures the inherent indirect upstream reciprocity behavior\cite{greiner2005indirect} of the user where prior help from a robot motivates the user to help another robot to increase their wellbeing. Finally, the policy also captures how to quantify the user's braking behavior to estimate the scooter's optimal action. 

Comparing individuals' social interaction perception across Figures \ref{fig:policy_socContent2}, \ref{fig:policy_socContent4}, and \ref{fig:policy_socContent6}, we see that higher individuals' content with their relationship with sidewalk and road users (Social interaction perception: content), they are more inclined toward not yielding to the robot even for higher levels of wellbeing. However, comparing individuals'  perception of their wellbeing with experiences with sidewalk and road users (Social interaction perception: ego wellbeing) across Figures \ref{fig:policy_socEgoWb2}, \ref{fig:policy_socEgoWb4}, and \ref{fig:policy_socEgoWb6}, we see that individuals with higher response values are more inclined to yield to the robot for their wellbeing to increase. This is expected as more pro-social individuals prefer to help the robot by yielding.  This demonstrates that the learned policy can distinguish users' perceptions of social interactions to ensure optimal scooter actions.

In summary, we show that the users' self-reported survey was influenced by their interaction with others and their intent. We observe that any yielding behavior contributed toward the ego would lead to enhancing their wellbeing and trust. However, users' social interaction perception can impact the influence of other factors on the users' change in wellbeing. We demonstrated that a  policy based on a simple machine learning model could quantitatively capture optimal actions by an AV scooter to enhance users' wellbeing. Future work will evaluate these policies using user studies to capture their efficacy. Finally, we want to acknowledge some limitations of the study. Given that our observational user study used pre-recorded videos and the users did not have active control over the vehicle, the lack of sense of control in AVs can affect wellbeing and trust \cite{welzel2010agency}. To mitigate this, we requested participants to indicate their intent to accelerate and decelerate throughout the study. Moreover, the proposed model is limited to a dyadic interaction between a self-driving scooter as an AV and a delivery robot as others. Additionally, maneuvers beyond yielding and non-yielding behaviors in future studies can help understand different types of ego and other agent actions. Given the limited scenarios and data sample size, complex machine-learning models were not used in this work. A larger sample size can allow training complex models that consider not only other factors but also dynamic relations across time to improve model performance. Furthermore, multi-objective policies can be calculated to consider other cognitive states like user trust and workload along with goal-related costs.
Nonetheless, the presented work demonstrates a significant step toward human-aware automation that can enhance wellbeing in mobility through optimal policies.



\section{Conclusion}
The recent emergence of automated vehicle (AV) technology and shared mobility has presented several challenges for how shared mobility systems like electric scooters are likely to influence the wellbeing of other road users. A key challenge is to understand the nature of these interactions and how prosocial interactions between agents may impact scooter users' wellbeing, trust, and satisfaction. We conduct an online video-based user study to evaluate users' subjective ratings of wellbeing, trust, satisfaction, and relationship with other road users during interactions between a self-driving scooter and delivery robots. We found that the yielding behavior of the delivery robot and AV scooter influenced the well-being, satisfaction, and trust of other road users. 
Based on the findings, we show that a prediction model for change in user wellbeing can be used to identify optimal actions of the AV scooter that increases wellbeing. The learned policy shows that the optimal actions are not only dependent on the robot and user's behavior but also depend on users' predispositions about their social interactions. The findings from this study provide a step toward designing personalized AV policies that could aid in ensuring optimal actions and outcomes in a shared environment to enhance users' wellbeing.  

\section{Acknowledgment}
We acknowledge the support from Hiu-Chun Lo at  Honda Research Institute USA., Inc for designing and implementing the web platform.

\bibliographystyle{IEEEtran}
\bibliography{main.bib}


\end{document}